# Education and Food Consumption Patterns: Quasi-Experimental Evidence from Indonesia

Dr Mohammad Rafiqul Islam[a,1] and Dr Nicholas Sim[b,2]

[a] Department of Economics, Shahjalal University of Science and Technology, Bangladesh.

[b] School of Business, Singapore University of Social Sciences

E-mail: rafiqieco@gmail.com

[1] Corresponding Author, Professor of Economics, Shahjalal University of Science and Technology, Sylhet 3114, Bangladesh. Email: rafiqieco@gmail.com
[2] Associate Professor, and Head, Graduate Programmes in Analytics and Visualisation, School of Business, Singapore University of Social Sciences, 463 Clementi Road, Singapore 599494

# Education and Food Consumption Patterns: Quasi-Experimental Evidence from Indonesia

*ABSTRACT: How does food consumption improve educational outcomes is an important policy issue for developing countries. Applying the Indonesian Family Life Survey (IFLS) 2014, we estimate the returns of food consumption to education and investigate if more educated individuals tend to consume healthier bundles than less-educated individuals do. We implement the Expected Outcome Methodology, which is similar to Average Treatment on The Treated (ATT) conceptualized by Angrist and Pischke (2009). We find that education tends to tilt consumption towards healthier foods. Specifically, individuals with upper secondary or higher levels of education, on average, consume 31.5% more healthy foods than those with lower secondary education or lower levels of education. With respect to unhealthy food consumption, more highly-educated individuals, on average, consume 22.8% less unhealthy food than less-educated individuals. This suggests that education can increase the inequality in the consumption of healthy food bundles. Our study suggests that it is important to design policies to expand education for all for at least up to higher secondary level in the context of Indonesia. Our finding also speaks to the link between food-health gradient and human capital formation for a developing country such as Indonesia.*

*Key Words: Food Consumption Parameters, Average Treatment on the Treated (ATT), Healthy Food and Unhealthy Food*

## 1. Introduction and background

The positive association between education and living standards is a well-established fact in social sciences (Case, 2006, Duflo, 2001, Psacharopoulos, 1994, Case and Deaton, 1999, Lucas, 1988). However, what are the channels for this association are still under debate. In recent times, there has been an increasing interest to investigate how education may affect choices in food consumption. The focus on food consumption is important since in developing countries, food is closely related to welfare and poverty (Todaro and Smith, 2012 and Goulet, 1997). Furthermore, studying the impact of education on food consumption may help uncovering food behaviour to design national nutritional policies. In a recent study, Wantchekon, Klasnja and Novta (2015) have noted that education can have an intense transformational impact on individuals and communes. Another study by Moreira and Padrao (2004) finds that education is one of the key elements to influence shifting consumption towards healthy food groups like vegetables and fruits. However, these studies have not ascertained whether the relationship is causal.

While the monetary returns to education (for instance, impact of education on earnings) in both developing and developed countries including Indonesia's, are well documented, causal studies on nonmonetary returns to education (for example, food consumption) are scarce. In this chapter, we examine the impact that education may have on food consumption. In particular, we test if more educated household heads tend to make a healthy food choice or unhealthy food choice. The findings may have policy implications for developing countries to invest more in education if it can be established that better educated individuals would choose to consume healthier foods.

Indonesia has made great advances in many areas including larger investment in the education and infrastructural development. Indonesia's expenditure on education as a share of GDP is about 3.4 percent and it has 50 million students and 250,000 schools (World Bank, 2014). Currently, the country has 9 years of compulsory schooling (CIA World Fact Book, 2014). Moreover, Indonesia's enrolment rates both at primary and secondary levels have been increased dramatically over the last few decades.[3]

There are several reasons why we focus on the impact that education has on food consumption. The first reason, due to survival and existence, food consumption is fundamental for the existence of life regardless of the level of education attained by an individual. Hence, from the policy standpoint on welfare, consumption is more relevant than earnings as labour income would ultimately translate into consumption.

Because lifetime consumption is smoother than income, consumption resembles more like a log normal distribution than income itself. Battistin, Blundell, and Lewbel (2009) observe that consumption expenditures across households are more log normally distributed while significant departures from log normality is found in income data. They note that consumption within a cohort has several implications for welfare and econometric modelling. Likewise, income data are noisier than consumption. It is anticipated that education has both lifetime productive returns and have had more role on consumption smoothing that can serve as a better measure for welfare than other educational outcomes, such as permanent income and earnings (Fulford, 2014).

Moreover, it is possible that a person does not have any earnings in the short run. For instance, a labourer who works in the agricultural sector sometimes suffers in seasonal unemployment and ends up with no earnings. However, he has to consume regularly, which makes consumption proportional to lifetime resources and reflects living standard throughout the year (see, Deaton and Grosh, 1998; Musgrove, 1978 and 1979; Paxson, 1992 and 1993; Wolpin, 1982).[4]

The second reason is that it is plausible that higher carbohydrate consumption due to lack of proper nutrition knowledge may result in health-risk; for example, among the poor segments in developing countries, higher price shock may force them to live below the required amount of calories (Abdulai and Aubert, 2004).

The final reason is that everyone in the households consume although not all members earn and have similar levels of education. Head of the household invests in children and schools in the expectation of accruing potential income in the household at the cost of current postponed adult consumption. As such, education has a prospective consumption returns.

Applying IFLS 2014 data to a semi-parametric model, this study finds that individuals who have received upper secondary school or higher levels of education, on average, consume 31.5 percentage points more healthy foods than those who have lower secondary school education or less. In terms of unhealthy food consumption, more-educated individuals, on average, consume 22.8% less unhealthy foods than less-educated individuals.

This study contributes to the food consumption literatures in two ways. First, this study is unique in the sense that it attempts to provide the first quasi-experimental evidence of the impact of education on food consumption patterns in Indonesia, while the literature has mainly focused on the impact of education on earnings or health or schooling for the next generation. In particular, this study investigates the role of education on choosing healthy food group or unhealthy food group by exploiting an exogenous variation of schooling-time required to attend school-to construct an IV. Second, it adds to the literature on the consumption returns to education by attempting to estimate the causal effect of education, while the literature has mainly focused on the correlation between education and food consumption (see, for example, Michael, 1975; Fulford, 2014; Bhandari, 2008; and Alem and Soderbom, 2012).

---

[3] Since the 1970's both the primary and secondary enrolment rates have increased dramatically in Indonesia (Economist, 2014). One reason for this huge increase in enrolment has been identified by Duflo (2001) that between 1973 and 1978 more than 61,000 primary schools built in Indonesia under the major school construction program, the Sekolah Dasar INPRES program.[3] About ten years later (initiated in 1973) government implemented compulsory education for primary school children (7-12 years). Consequently, primary school participation rate rose to 92 percent in 1993 compared to 79 percent 10 years before.[3] Again in 1994, the country expanded compulsory education to 9 years for every Indonesian in the 7-15 age group. Since 2009, the government has allocated one-fifth of its yearly budget in education.

[4] A remarkable quotation about 200 years back by Anthelme Brillat -Savarin noted in Anand and Sen (1998), "Tell me what you eat," and I will tell you what you are."

## 2. Literature review

Three interrelated groups of literatures are dominant in explaining the impact of education on sociol-economic outcome: earnings, health, and growth. A vast number of literatures investigate the relationship between education and earnings including compulsory schooling and earnings ((Angrist and Krueger, 1991; Stephens and Yang, 2014), returns to schooling from sibling data (Ashenfelter and Krueger, 1994; Butcher and Anne, 1994 and schooling and selectivity bias (Garen, 1984, education, ability and earnings)).

Koc and Kippersluis (2015) investigated Discrete-Choice-Experiment (DCE) of educational disparities on making food consumption and found that health knowledge differentials play a greater role in education disparity in food in Netherlands. Cutler and Lleras-Muney (2010) find that there is a strong disparity in healthy behaviours like diet choice across education groups. Haines, Guilkey and Popkin (1988) examined the food consumption decisions as a two-step process decomposing food groups into low fat milk versus high fat milk and high fat low fibre bread group. Their findings suggest that decision to consume a specific food within a food group is statistically significantly different from how much to consume for more broadly defined food groups.

Fulford (2014) examined the returns to education in India and found that an additional year of education brought 4 percent more consumption of male cohorts with no extra consumption for female cohorts. A related study in Ethiopia by Alem and Soderbom (2012) found that a significant percentage of households adjusted food consumption due to large price shock. Yen, Lin, and Davis (2008) explored the linkage between consumer knowledge and meat consumption both at home and away from home and found that dietary knowledge reduced beef and pork consumption both at home and away from home and men consumed more meat and fish than women. Abdulai and Aubert (2004) conducted parametric and nonparametric analysis of calorie consumption in the presence of behavioural heterogeneity and measurement error using panel data from Tanzania and concluded that higher food prices could reduce calorie demand significantly, and as such, it would be important to allocate targeted food subsidies for poor households. Cain et al. (2010) conducted an empirical study using household level consumption expenditure data from India and concluded that the amount of inequality generated by the education of household heads were much greater than the sum of all other household characteristics. In particular, they found that education accounted for 8% and 9% of inequality in 1993 and 2004 in rural areas and the respective figures were 25% and 28% in urban areas in the same years.

Studies also found that consumption expenditures were strongly correlated with education. In a recent study on Nepal, Fafchamps and Shilpi (2014) have found that there is a strong statistical association between male education and household's welfare even after controlling educational attainment within their birth cohort. In another benchmark study, Michael (1975) has found that the education elasticity of goods is -0.07 and of services is 0.19 meaning that an additional year of schooling shifts the spending patterns toward services. A number of experimental studies have linked diseases to the choices of food consumption. Therefore, health and development practitioners have been concerned about understanding which factors could influence consumption patterns of the population (Alam and Hossain, 2018, Fraser et al. 2000, Chait et al. 1993, Potter, 1997, and Denke, 1997).

Ricchiuto, Tarasuk, and Yatchew (2006) characterized the role of household's socioeconomic status in choosing a particular food in Canadian households. They concluded that irrespective of household size, income and composition, higher education was associated with the purchase of larger quantities of vegetables, milk, and food products. Interestingly, households with post-secondary education purchased 6% more fruit and vegetables than those who have fewer than 9 years of schooling. Some other studies also obtained similar findings that higher income and higher education are associated with consuming more vegetables and fruits (Alam, 2013, Nayga et al., 1999; Groth et al., 2001, Perez, 2002).

Duflo (2001) found that the economic returns (i.e., earnings) from an additional year of education in Indonesia ranged from 6.8% to 10.6%. Although she generated huge variations of schooling by exogenous district-level changes of the number of schools constructed by the Indonesian government, other local factors (for instance, district level teacher-student ratio) could threaten the exclusionary restriction of her instrument. Purnastuti, Salim, and Joarder (2015) examined returns to schooling in Indonesia using IFLS 2007 and an IV approach. Their OLS estimates show that the returns to schooling is 4.36 percent for males and 5.26 percent for females. However, the relationship between education and earnings is not significant when IV is used. A similar work by Comola and Mello (2010) for the Indonesian labor market had shown that the returns to education from 9.49% to 10.32%, although their work did not address the identification issues that are standard when estimating the returns to education. Dumauli (2015) examined the private returns to education in Indonesia accounting for sample selection and endogeneity issues. The household FE estimates indicate the returns to education fell from 10.8% to 5% between 1986 and 2007. This may be a reason for why college enrolment rate in Indonesia has stagnated during this period.

The theoretical foundations of schooling as a formation of human capital and its impact on monetary returns are quite strong in the literatures. However, the impacts of education on non-monetary returns are scarce in the empirical literatures, although studying education's impact on non-pecuniary outcomes is indispensable at its own right. This paper exploits quasi-natural experiment to estimate the food consumption returns to education in Indonesia which is the main departure from the existing studies that investigate the returns to education.

## 3. Conceptual framework of linking education to consumption

There are a number of ways that education might affect food consumption. First, education may enhance the capacity of understanding of nutritional aspects of variety of foods that may lead to consume the healthy foods. Second, more highly educated people may have higher earnings potentials, which may lead to greater access to varied food groups in the market. Third, higher educated individuals may spend more time on electronic media and newspaper that have a coverage of food and nutrition and thus may have more knowledge of how unhealthy food choices would be threatening to health. Fourth, more highly educated individuals may build up healthier dietary habits more quickly than lower educated individuals. Fifth, they may have broader cultural views of food consumption that lead to diversity of food consumption. Finally, compared to other demographic factors such as age and gender, education is a policy variable that is responsive to government interventions. If government is concerned with the health aspects of individuals, it may introduce nutrition education in the schools.

## 4. Data and descriptive statistics

The empirical analysis in this paper draws on the publicly available household level panel datasets of IFLS. we use data from the fifth wave of the IFLS fielded from September 2014 to March 2015. The IFLS 2014 is an ideal data choice for our case as Indonesia has been gradually transforming into decentralized economy followed by deregulation while government has been allocating more finance to expand since 2000. In particular, implementation of compulsory education policies in 1974 and 1984 approved by the government are likely to facilitate cohort-specific individuals to complete full time education. For a detailed description of the IFLS 2014 survey see Strauss, Witoelar, and Sikoki (2016).

     IFLS collects a wide range of information at the individual, household and community level. The IFLS sample is drawn from 321 randomly selected villages covering 13 Indonesian provinces and representing 83% of the country's population. The last survey is carried out in 2014. The sub-sample I use consists of household head aged 15-70 and those who have reported non-missing food consumption and schooling information. The dependent variables in our analysis are: log of per capita healthy food consumption and log of per capita unhealthy food consumption at the household level.[5] The final sample contains about 13000 households.

     Table 1 presents descriptive statistics for the main variables used in this study. It shows that individuals with higher secondary or more levels of education have, on average, 0.33 log points higher than those with less than higher secondary education. They have 8.23 extra years of schooling. Graduates from the higher secondary schools or more are likely to come from families with better educated parents and have fewer household members in the family. Higher secondary or more educated individuals are more likely to live in urban areas than rural areas and are less likely to consume unhealthy foods. They also tend to live in the proximity of high schools than lower educated persons.

---

[5]Following Usfar and Fahmida (2011), we have constructed healthy and unhealthy food groups using IFLS consumption module: i) the main staples, vegetables and fruits, meat and animal products, and fish constitute a healthy food group; and ii) the dried foods, condiments, and other foods constitute an unhealthy food group. Both food groups have been converted into annualized per capita food group at the household level.

**Table 1**: Summary statistics for the treatment and control groups

|  | Higher secondary or more (Treatment group) | Less than higher secondary (Control group) |
|---|---|---|
|  | *N* = 9945 | *N* = 4978 |
| Log of per capita healthy food consumption | 14.557 (0.872) | 14.231 (0.861) |
| Log of per capita unhealthy food consumption | 14.374 (0.966) | 14.892 (0.934) |
| Years of education | 14.787 (3.348) | 6.508 (1.856) |
| Household size | 5.046 (3.115) | 5.817 (3.135) |
| Age | 39.098 (11.866) | 44.632 (12.867) |
| Sex | 0.853 (0.353) | 0.822 (0.382) |
| Employment | 0.721 (0.448) | 0.683 (0.665) |
| Married | 0.863 (0.343) | 0.955 (0.206) |
| Muslim | 0.852 (0.354) | 0.914 (0.278) |
| Catholic | 0.021 (0.144) | 0.009 (0.095) |
| Protestant | 0.059 (0.235) | 0.037 (0.188) |
| Other | 0.066 (0.249) | 0.038 (0.193) |
| Javanese | 0.392 (0.488) | 0.465 (0.498) |
| Sundanese | 0.108 (0.311) | 0.136 (0.343) |
| Minang | 0.065 (0.247) | 0.045 (0.208) |
| Other | 0.434 (0.495) | 0.352 (0.477) |
| Fathers education | 6.123 (2.067) | 5.202 (2.425) |
| Mothers education | 4.112 (2.101) | 4.011 (2.001) |
| Distance to school (minutes) | 16.145 (12.185) | 16.321 (12.079) |
| Distance to health post (km) | 5.009 (8.357) | 6.183 (10.461) |
| Rural household | 0.249 (0.432) | 0.462 (0.498) |
| North Sumatra | 0.080 (0.271) | 0.074 (0.261) |
| West Sumatra | 0.046 (0.210) | 0.041 (0.198) |
| South Sumatra | 0.048 (0.214) | 0.046 (0.210) |
| Lampung | 0.028 (0.166) | 0.045 (0.208) |
| Jakarta | 0.078 (0.269) | 0.059 (0.235) |
| Central Java | 0.090 (0.286) | 0.136 (0.342) |
| Yogyakarta | 0.066 (0.249) | 0.040 (0.196) |
| East Java | 0.108 (0.310) | 0.149 (0.356) |
| Bali | 0.063 (0.244) | 0.039 (0.194) |
| West Nusa Tenggara | 0.088 (0.284) | 0.062 (0.241) |
| South Kalimantan | 0.043 (0.204) | 0.044 (0.205) |
| South Sulawesi | 0.047 (0.212) | 0.048 (0.214) |
| Rural | 0.249 (0.432) | 0.462 (0.498) |

*Source:* Calculated from the IFLS 2014 and sample is restricted to the non-missing schooling and distance to the school.

## 5. Methodology

The estimation approach in this paper is carried out in three steps. First, we set up the model of consumption returns to education. Second, we elucidate the endogeneity issues of education. Third, we explain of building up the scenario of IV.

*5.1 A semiparametric selection model*

Estimating marginal food consumption returns to education is a key parameter of interest in this study. In other words, estimation of the food consumption returns to education is one of the central focuses for the policy makers to evaluate cost and benefit of the policy e.g., educational expansion policy of the government (Carneiro, Heckman, and Vytlacil 2011). The figure 1 exhibits density of returns to education intuitively.

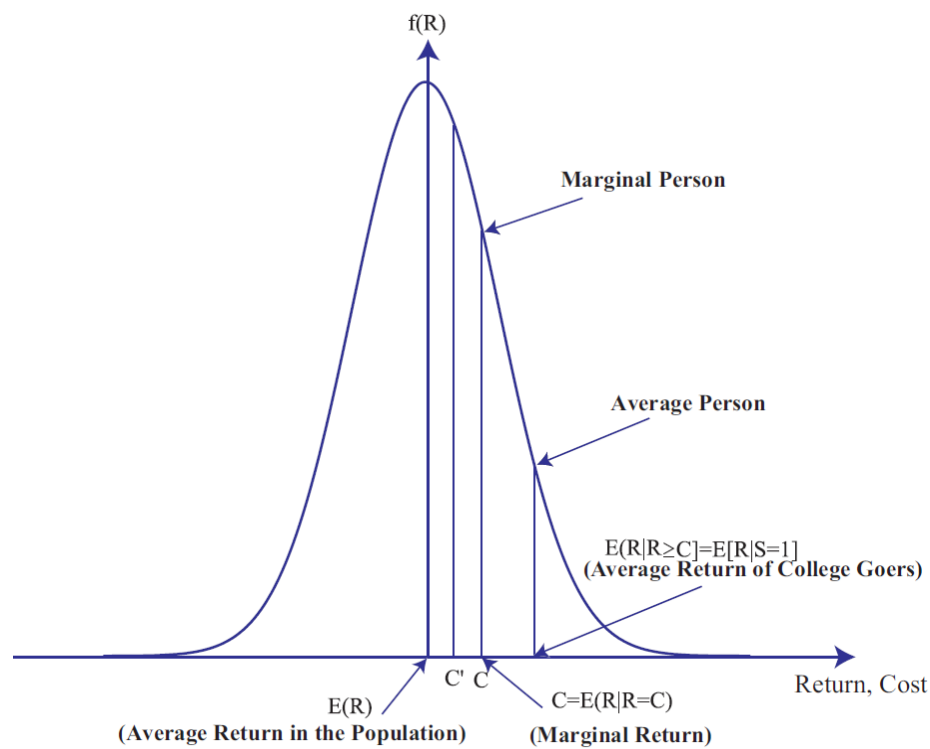

**Figure 1:** Density of Returns to Education. *Source:* Carneiro, Heckman and Vytlacil (2003)

The mean marginal returns to education can be estimated by the following equation:

$$\ln Y = \alpha + \beta S + \varepsilon \qquad (1)$$

where *lnY* is the log of per capita food consumption, *S* is a dummy variable indicating if an individual (i.e., household head) has had a high school education, β is the returns to schooling of the household head (which may differ among individuals), and ε is the residual. The coefficient *β* would be positive if an individual chooses healthy food groups and it would be negative if he chooses an unhealthy food group. It is assumed here that knowledge about diet-health relationship induces what to consume. The effect of graduating from high school on food consumption may be confounded by self-selection. I will address this issue by using an IV or implementing model that corrects for selection.

We follow Carneiro, Lokshin, and Umapathi's (2016) model of potential outcomes applied to education. Consider a model with two levels of schooling:

$$\ln Y_1^{(k)} = \alpha_1 + X\beta_1 + U_1 \qquad (2)$$

$$\ln Y_0^{(k)} = \alpha_0 + X\beta_0 + U_0 \qquad (3)$$

$$S = 1 \text{ if } Z\lambda - U_s > 0 \qquad (4)$$

$\ln Y_1^{(k)}$ is the log of per capita food consumption with $k$ equal to healthy or unhealthy food consumption if the head of the household have completed higher secondary education or more, $\ln Y_0^{(k)}$ is the log or per capita food consumption if the head of the household have completed less than secondary education, $X$ is a vector of observable characteristics which affect food consumption, and $U_1$ and $U_0$ are the error terms, $Z$ is the other vector of household characteristics affecting schooling. From now onwards, we drop superscript $k$ for convenience.
Equation 4 can be rewritten as:

$$S = 1 \text{ if } P(Z) > V \qquad (5)$$

where $P(Z) = F_{U_s}(Z\lambda)$ and $V = F_{U_s}(U_s)$ and $F_{U_s}$ is a cumulative distribution function of $U_s$. $V$ is assumed to be uniformly distributed.
The observed food consumption can be written as:

$$\ln Y = S \ln Y_1 + (1 - S) \ln Y_0 \qquad (6)$$

Food consumption returns to schooling can be expressed as:

$$\ln Y_1 - \ln Y_0 = \alpha_1 - \alpha_0 + X(\beta_1 - \beta_0) + U_1 - U_0 \qquad (7)$$

Note from Equation 7 that consumption returns to education vary across individuals with different *X*'s and different *U₁* and *U₂*. This implies heterogeneity in returns to education and the Equation 7 also aids to make a distinction between the returns for average and marginal individuals.
The marginal treatment effect (MTE) is the main object to be estimated here. The MTE can be expressed as follows:

$$MTE(x,v) = E(\ln Y_1 - \ln Y_0 | X = x, V = v) = \alpha_1 - \alpha_0 + x(\beta_1 - \beta_0) + E(U_1 - U_0 | X = x, V = v) \qquad (8)$$

The MTE measures the consumption returns to education for individuals with different levels of observables ($X$) and unobservables ($V$). Hence, MTE offers a characterization of heterogeneity in returns. $(\beta_1 - \beta_0)$ could be positive or negative. If parents were highly educated, children would be more educated as well. In this case, we expect higher consumption returns to schooling, which implies that $(\beta_1 - \beta_0) > 0$. If parents were less educated, children would be less educated as well. In this case, we expect lower consumption returns to education. Likewise, $V$ may be interpreted as positive unobserved ability. Individuals with high values of $V$ are less likely to enrol in school than those with low values of *V*. If this is the case, $E(U_1 - U_0|X = x, V = v)$ would tell us how the returns to education vary with unobserved ability.

Several other parameters of interest can be constructed as weighted averages of the MTE. Heckman and Vytlacil (2005) show that the following important parameters can be derived from MTE:

$$ATE(x) = \int MTE(x,v) \int f_{v|x}(v|x)dv \tag{9}$$

$$ATT(x) = \int MTE(x,v) \int f_{v|x}(v|x, S = 1)dv \tag{10}$$

$$ATU(x) = \int MTE(x,v) \int f_{v|x}(v|x, S = 0)dv \tag{11}$$

where *ATE(x)* is the average treatment effect, *ATT(x)* is average treatment on the treated, *ATU(x)* is average treatment on the untreated (conditional on *X=x*, and $f_{v|x}(v|x)$ is the density of *V* conditional on *X*.

Another policy related parameter of interest can be constructed as follows:

$$PRTE(x) = \int MTE(x,v) \int f_{v|x}, (v|x, S(z) = 0, S(z') = 1)dv \tag{12}$$

where PRTE(x) is the policy related treatment effect which measures the average returns to schooling for those induced to change their schooling decision in response to a specific policy assuming policy shifts *Z* from *Z=z* to *Z=z'*.

To estimate MTE, we use the method of local IV that imposes no distributional assumptions on the unobservables of the model apart from the assumptions that X and Z are independent of error terms.

There are two steps in the procedure for estimating MTE. The first step is to estimate a regression of the outcome *lnY*, on *X* and *P* (propensity scores). This step can be written as follows:

$$\begin{aligned} E(lnY|X,P) &= E[\alpha_0 + X\beta_0 + S(\alpha_1 - \alpha_0) + SX(\beta_1 - \beta_0) + U_0 + S(U_1 - U_0)|X,P \\ &= \alpha_0 + X\beta_0 + P(\alpha_1 - \alpha_0) + PX(\beta_1 - \beta_0) + E(U_1 - U_0|S = 1, X, P)P \\ &= \alpha_0 + X\beta_0 + P(\alpha_1 - \alpha_0) + PX(\beta_1 - \beta_0) + K(P) \end{aligned} \tag{13}$$

K(P) is a flexible function of P. We will estimate it using a non-parametric method such as local linear regression. The regression in (19) is partially linear, where *X* and *XP* are partially linear, and the function *K (P)* is non-parametrically estimated.

The second step is to take derivative of (9) with respect to *P* to get MTE:

$$MTE(x,v) = \frac{\delta E(Y|X,P)}{\delta P} = X(\beta_1 - \beta_0) + K'(P) \tag{14}$$

Therefore, the local IV estimator for the Equation 1 and 2 requires regressing *ln(Y)* on *X* and *P* and taking the partial derivative of the estimated regression function with respect to *P*.

*5.2 Endogeneity of education*

If the error term in Equation 1 (the food consumption equation) were orthogonal to the regressors, the OLS estimates of the parameters in Equation 1 would be unbiased and consistent. However, these OLS estimates, which capture the impact of education on food consumption, may still be biased, particularly in the case of self-reported schooling. It is noted in paper by Lleras-Muney (2005) that the OLS estimates of education may be biased due to measurement error in education. Likewise, Angrist and Krueger (1999) conclude that with no controls, measurement error can shrink returns to education by about 10 percent. Griliches (1997, 1979) underscores that measurement errors in education would lead a downward bias in the estimates of schooling on earnings. Card (1999) points out that errors in reported schooling may be mean-regressive as higher educated individuals cannot state positive errors in schooling and lower educated individuals cannot state negative errors in schooling.

Equation 1 may also suffer from endogeneity problem due to omitted variable. For example, ability or current health condition of a household head may be correlated with both education and food consumption. Or, there may arise endogeneity due to reverse causation. For instance, healthy food choices may increase life expectancy which would lead to persuade to take schooling at a longer time.

*5.3 Validity of the instrumental variable*

We deal with the endogeneity of education by employing IV and MTEs. To do so, years of schooling is instrumented with the time required (measured in minutes) to attend school from home (distance to the school). Distance to the nearest school as an instrument is first used by Card (1993) and successively has been used by other researchers (Kane and Rouse, 1995, Kling, 2001, Currie and Moretti, 2003, Cameron and Taber, 2004, Carnerio, Heckman and Vytlacil, 2011, and Carnerio, Lokhsin, Ridao-Cano, and Umapathi, 2016). When compulsory education laws come into effect (e.g. Compulsory 6 years schooling law in 1974 and compulsory 9 years schooling law in 1984 in Indonesia), the decision to take further education is less about tuition and more about location. This is particularly true for developing countries where there is inadequate infrastructure, transportation, and schools.

In order for distance to the school to be a valid instrument, it needs to satisfy two conditions: i) it should have a strong correlation with years of schooling and ii) it should have no direct linkage with outcome variable (food consumption per capita). In this connection, two main concerns need to be mentioned. First, households and schools may not be located randomly across localities in Indonesia. Second, although distance is measured while a student completed education, distance is not measured at the time of schooling decision. As it is measured at the time of survey year, there is likely reporting error in minutes or it is an approximation like Card (1995). The main problem with this sort of approximation is that educated families or individuals may have already shifted to the place where there are more private and public schools and good infrastructures. This could entail reverse causality in the first stage relationship.

A consistent estimate of the impact of education on food consumption can be obtained if a variable that affects education but not food consumption can be formed. In this case, one needs to identify a causal determinant of education that can be legitimately excluded from the food consumption equation. Distance from the community to the nearest secondary schools may be used as a valid instrument to education as it presumably affects food consumption only through education, and it may not influence other household level determinants of food consumption. For instance, individuals who grow up in an area without secondary schools face a higher cost of attending schools. This higher cost reduces investments in further education, particularly for the children with low-income families. This implies that growing up near secondary schools may have a larger impact on the education outcomes at least in terms of enrolment, particularly for the children from economically disadvantaged families.

There are some reasons why individuals who grow up in close proximity to schools may have higher consumption than others. First, Families that put a strong emphasis on education may choose to reside near schools. Children of such family background may have higher earning potentials the labour market that would translate into higher consumption. Second, the presence of schools at nearby communities reduces transaction cost, largely due to easy driving distance to attend in the schools. Hence, families have some scopes to allocate more on consumption. Finally, it is possible that nearby schools are also located in close proximity to industries and markets. Hence, both higher labour market earnings potentialities and greater access to goods may result in higher consumption. So, all of the points imply that distance to schools from the community is likely to serve as a valid instrument to education.

Table 2 indicates that distance to the nearby higher secondary school in the community is a strong predictor of schooling enrolment in the higher secondary level. we run a logit regression where the dependent variable is a dummy variable that indicates (i.e., is equal to 1) if the individual has attended higher secondary schools or more. we control for age, employment and marital status of the household head, religions, ethnicities, parental education, distance to the nearest school, and distance to the nearest health post as a proxy for location geographies, and rural dummy as a proxy for area.

**Table 2:** Higher secondary school decision model – average marginal derivative

|  | (1) Coefficients | (2) Average Derivative |
|---|---|---|
| Distance to school in minutes | -0.223*** | -0.020*** |
|  | (0.030) | (0.006) |
| Age | 0.036** | -0.015** |
|  | (0.002) | (0.001) |
| Employment | 0.172 | 0.172 |
|  | (0.164) | (0.164) |
| Married | -1.312* | -0.019* |
|  | (0.744) | (0.744) |
| Protestant | 0.640*** | 0.042*** |
|  | (0.157) | (0.003) |
| Catholic | 0.914* | 0.014* |
|  | (0.503) | (0.008) |
| Other religions | 0.618* | 0.618* |
|  | (0.330) | (0.330) |
| Javanese | -0.567*** | -0.567*** |
|  | (0.172) | (0.172) |
| Sundanese | -0.762*** | -0.762*** |
|  | (0.253) | (0.253) |
| Minang | 0.302 | 0.302 |
|  | (0.352) | (0.157) |
| Father higher education | 0.824*** | 0.231*** |
|  | (0.040) | (0.003) |
| Mother higher education | 0.441*** | 0.154** |
|  | (0.003) | (0.127) |
| Rural | -0.918*** | -0.092*** |
| | (0.168) | (0.101) |
| Distance to health post in km | -0.019 | -0.025 |
| | (0.008) | (0.005) |
| Constant | 1.256 | 1.112 |
| | (0.836) | (0.225) |
| Location fixed effect | | yes |
| Test for joint significance of instruments: Chi-square/p-value | | 11.63/0.0011 |

*Notes:* This table presents the coefficients and average marginal derivatives from a logit regression of higher secondary school attendance (an indicator variable that is equal to 1 if an individual has ever attended higher secondary school or more and equal to 0 if he has never attended higher secondary school but graduated from lower secondary school) on household level observables. Type of location is controlled for using province dummy variables.
***significant at 1 percent level; **significant at 5 percent level and; *significant at 10 percent level.

## 6. Empirical results

*6.1 Standard estimates of the food consumption returns to education*

In the empirical model, we divide schooling into two categories: i) completed lower secondary or below, and ii) completion of upper secondary or higher. While this division bands together different levels of schooling, it simplifies the model and has been considered in the literature (e.g., Willis and Rosen, 1979).[6] The transition to upper secondary schooling is of interest in the Indonesian context given its current effort to expand secondary education. Higher secondary schooling corresponds to 10 or more years of completed education in Indonesia.

We use the distance (in minutes) from the household to the nearest secondary school in the community as an instrument for schooling. The distance is self-reported by the respondent who is an adult, 15 years or older, and the distance in recorded in section DL in Book IIIA in the household survey.

There are two endogeneity concerns that come along with using the distance of the household to the school as an instrument. First, the distance between school and house can be a choice as households may locate themselves closer to school in order to make it easier for their children to go to school. Second, distance may be related to wellbeing of the household if schools are centrally located and poorer households are living close to central school locations. There are three ways I have dealt with these sources of simultaneity. First, the location of the school may be exogenous when a set of controls are rich enough to capture those sources of endogeneity. we have controlled a detailed set of household and regional characteristics to minimize reverse causality. Specifically, we have used distance to the nearest health post and provincial dummies as a proxy for location characteristics and found that the distance to the nearest health post does not predict school choice. Second, the well-off households may choose to live in the close proximity of the schools than poorer households. To minimize this problem of endogeneity, we have elsewhere used log of household head earning as a control, and the impact of education on healthy food consumption is found to be statistically insignificant in an instrumental variable regression when log of earnings is added as a control. Third, nonetheless one could concern that instrument is weak. Hence, we have additionally used the estimated propensity score and the interaction of a set of controls and the distance as an instrumental variable to minimize the threat that the results are driven and confounded by the other household level unobserved factors.

---
[6] Other schooling might have taken as categorical variables such as no schooling, elementary, high school, college, and tertiary education. Impact of each level of schooling on food consumption at the household level is difficult to establish as no individual food consumption data are available in the surveys.

Table 3 presents results of standard OLS and IV estimates. The key variable of interest is schooling: takes 1 for household head who has completed higher secondary education or more, and takes 0 for less than higher secondary schooling. I use the log of per capita healthy food consumption and the log of per capita unhealthy food consumption as dependent variables. All specifications in Table 1.3 includes the following controls: household size, age of the household head in years, indicator variables for whether head of the household is employed and married, indicators for main religions (Protestant, Catholic, other religions and Muslim is an omitted category), indicators for main ethnicities (Javanese, Sundanese, Minang, and 'Other ethnicities' are omitted), and fathers and mothers education measured with the highest grade completed parental education, distance to the nearest health post in kilometres, an indicator for rural households.

It is observed that individuals with higher secondary school education consume 31.5% more healthy foods than the individuals with lower secondary school education. The corresponding IV estimate is 33%. On the other hand, the reduction of unhealthy food consumption due to upper secondary education from our OLS and IV estimates are 22.8% and 26.9% respectively. In both types of food consumption, IV estimate produces larger coefficient than OLS. It may be attributed to the measurement error in the data or the differences in the magnitude may depend on the choice of instrument.

**Table 3:** OLS and IV estimates of food consumption returns to higher secondary schooling

|  | (1) OLS Healthy food | (2) IV Healthy food | (3) OLS Unhealthy food | (4) IV Unhealthy food |
|---|---|---|---|---|
| Schooling | 0.315*** | 0.330*** | -0.228*** | -0.269** |
|  | (0.050) | (0.043) | (0.050) | (0.552) |
| Household size | -0.061*** | -0.054*** | -0.090*** | -0.088*** |
|  | (0.007) | (0.008) | (0.009) | (0.012) |
| Age | -0.004 | -0.007 | -0.012*** | -0.004 |
|  | (0.003) | (0.004) | (0.003) | (0.005) |
| Employment | 0.192*** | 0.203*** | 0.210*** | 0.163** |
|  | (0.040) | (0.052) | (0.044) | (0.068) |
| Married | 0.645* | 0.210 | -0.777*** | -0.523 |
|  | (0.334) | (0.504) | (0.183) | (0.330) |
| Protestant | 0.108*** | 0.172** | -0.377*** | -0.115** |
|  | (0.026) | (0.055) | (0.132) | (0.051) |
| Catholic | -0.072 | 0.038 | -0.381** | 0.044 |
|  | (0.208) | (0.258) | (0.190) | (0.254) |
| Other religions | -0.125 | 0.060 | -0.515*** | 0.036 |
|  | (0.248) | (0.310) | (0.196) | (0.283) |
| Javanese | 0.040*** | 0.003*** | -0.014** | -0.030** |
|  | (0.010) | (0.001) | (0.007) | (0.011) |
| Sundanese | -0.063 | -0.081 | 0.043 | 0.298* |
|  | (0.099) | (0.157) | (0.092) | (0.175) |
| Minang | 0.091 | 0.116 | -0.056 | -0.437 |
|  | (0.217) | (0.297) | (0.149) | (0.302) |
| Fathers education | 0.017*** | 0.015*** | -0.014*** | -0.020*** |
|  | (0.002) | (0.007) | (0.001) | (0.003) |
| Mothers education | 0.027*** | 0.018*** | -0.015*** | -0.023*** |
|  | (0.002) | (0.008) | (0.002) | (0.007) |
| Rural | 0.011 | 0.459 | -0.264*** | 1.024 |
|  | (0.048) | (0.371) | (0.053) | (1.055) |
| Distance to health post (km) | -0.001 | 0.001 | 0.002 | 0.010 |
|  | (0.002) | (0.005) | (0.002) | (0.011) |
| Location controls | yes | yes | yes | yes |
| $F$-test of excluded instruments |  | 11.0 |  | 13.11 |
| Observation | 8000 | 8000 | 8000 | 8000 |
| R-squared | 0.167 | 0.158 | 0.301 | 0.37 |

Note: This table reports the coefficients for OLS and 2SLS IV for regression of log of per capita healthy food consumption (column I and column II) and log of per capita unhealthy food consumption (column III and column IV) on higher secondary education (higher secondary schooling (a dummy variable that is equal to 1 if an individual has completed upper secondary school and above and equal to 0 if he has completed schooling below upper secondary level).
***significant at 1 percent level; **significant at 5 percent level and; *significant at 10 percent level.

Some robustness checks have also been conducted and shown elsewhere[7]. The first one corresponds to the results presented in Table 3. In doing so, the definition of schooling (i.e., an indicator variable) is the same as defined in Table 1.3; however, log of household head annualized per capita earnings is added into the specification. The OLS coefficient of schooling for healthy food becomes smaller (0.198 versus 0.315) although highly statistically significant. The IV coefficient on schooling for healthy food becomes larger, but not statistically significant. It has happend due to the fact that IV estimates are very sensitive to the choice of instrument. On the other hand, the point estimates of OLS and IV (-0.136 and -0.153) from regressing unhealthy food consumption on schooling are much smaller, although highly significant, compared to the estimates (-0.228 and -0.269) presented in table 4.3. Another robustness check, is conducted, where log of earnings has been dropped and education is measured as years of completed education by the head of the household, and other controls. The signs of OLS estimates from both healthy and unhealthy food group consumption regressions are as expected. However, the magnitudes of the estimates are quite smaller compared to estimates in table C4.4 in both regressions. The signs of the corresponding IV coefficients are as expected. However, these coefficients are not statistically significant. The difference between OLS and IV coefficients may be attributed to measurement errors in schooling.

*6.2 Estimates of the average and marginal treatment effects*

Table 4 exhibits the average food consumption returns to higher secondary education for different groups of individuals. The return in terms of healthy food consumption to higher secondary school for a random person (ATE) is 10.2%. The return for the individuals who were enrolled in the higher secondary school (ATT) is marginally higher (11.3%). The return for the individuals who did not attend higher secondary school, had they been attended (ATU) is 7.1%. The average return for those induced to attend higher secondary school for a particular policy shift (PRTE) is 13%.[8] An estimate for the return to a marginal student (AMTE/MPRTE) is 12.3%.

**Table 4:** Estimates of average returns to higher secondary schooling (Dependent variable: log of per capita healthy food consumption)

| Parameter | Semi-Parametric Estimate | Normal Selection Model |
|---|---|---|
| ATT | 0.113 | 0.105 |
|  | (-0.003, 0.023) | (-0.004, 0.238) |
| ATE | 0.102 | 0.065 |
|  | (0.021, 0.254) | (-0.013, 0.122) |
| ATU | 0.701 | 0.017 |
|  | (-0.301, 0.831) | (-0.091, 0.176) |
| PRTE | 0.13 | 0.019 |
|  | (-0.046, 0.335) | (-0.018, 0.301) |
| MPRTE | 0.123 | 0.012 |
|  | (-0.057, 0.228) | (-0.019, 0.291) |

---

[7]Full results available upon request.
[8]The particular policy exercise we have executed is: a 15% reduction of a distance to higher secondary school and find the parameter to understand the impact of an education expansion program.

Note: This table records estimates of different consumption returns to higher secondary education for the semi-parametric and normal selection models: average treatment on the treated (ATT), average treatment effect (ATE), average treatment on the untreated, policy relevant treatment effect (PRTE), and the marginal policy relevant treatment effect (MPRTE). Bootstrapped highest posterior density 95% intervals are reported in parentheses.

*6.3 Additional sensitivity results*

Table 5 shows the additional sensitivity results of regressing education on healthy food consumption by employing IV using *P* (estimated propensity scores) and interactions of original IV (distance to the secondary school) and selected household observables to bring more variations education.

The OLS coefficient (0.172) of higher secondary education is positive and highly statistically significant (see Table 1.5). This implies that individuals who have attended higher secondary schools, on average, consume 17.2% more healthy food than those who have graduated from the lower secondary schools. When propensity scores (*P*) are used as an instrument to education, the coefficient is highly significant and very similar in magnitude to OLS estimate (0.189). When an interaction of distance to secondary schools with household observables (parental education, religion, ethnicities, and age) is used as instrument for education, the IV estimate gets quite large, though significant and expected sign is observed.

**Table 5:** Estimates of average returns to higher secondary schooling (Dependent variable: log of per capita healthy food consumption) - Additional sensitivity results

|  | OLS | IV ( $Z*X$ interactions) | IV using $P$ |
|---|---|---|---|
| Higher secondary education | 0.172*** | 0.253*** | 0.189*** |
|  | (0.022) | (0.025) | (0.049) |

Note: This table reports the coefficients for OLS and 2SLS IV for regression of log per capita healthy food consumption on schooling (an indicator variable that is equal to 1 if an individual has ever attended higher secondary school or more and equal to 0 if he has never attended higher secondary school but graduated from lower secondary school), controlling for parental education, religion and location. Column 1 shows the OLS results, controlling for parental education, religion, ethnicities, and age. Column 2 exhibits IV estimates and excluded instruments are distance to secondary school and interactions with parental education, religion, ethnicities, and age. Column 3 records the IV estimates and excluded instrument is the estimated propensity scores. Type of location is controlled using province dummies. Muslim is an omitted category for Muslim and other ethnicities for ethnicities. Standard errors are shown in parenthesis and are robust to clustering at the community level. All coefficients are significant at 1 % level.

**7. Conclusion**

Indonesia has been very successful for its initiative to expand education since 1970s. The enrolment rates are closely universal for elementary schooling and are about 75% for secondary education. Applying very recent data from Indonesia, this study explores the impact of education on food consumption. In particular, we have investigated whether education has a role to pick up consumption bundles, which have health implications.

We find that those who have completed higher secondary education or more substantially consume more healthy food and considerably reduce unhealthy food consumption. Specifically, individuals who have been graduated from upper secondary schools or higher educational institutions, on average, consume 31.5% higher healthy foods than those who have graduated from the lower secondary schools or less. With respect to unhealthy food consumption, more-educated individuals, on average, consume 22.8% less unhealthy foods than less-educated individuals. This implies a large inequality in consuming healthy food bundles where more education is a determinant. So, it is important to design policies to expand education at least up to higher secondary level for all in the context of Indonesia. This finding is important for better understanding of food-health gradient and human capital formation in a country like Indonesia.

However, one of the caveats of the above finding is that construction of healthy food group contains all staple foods and higher consumption of rice, which has the largest share in staple food, may not be always a worthy choice in respect to health and nutrition. Hence, it requires careful division of healthy and unhealthy food groups in the context of Indonesia. Without proper nutrition knowledge, the generalization of the result would be less practicable to analyse food consumption parameters in the determination of non-monetary returns to education. Furthermore, we have used IV when calculating the treatment effect, which may result in local treatment effect nonetheless.

## Acknowledgment:
We sincerely acknowledge the valuable comments by the renowned economists and social scientists for the presented paper titled "Education and Food Consumption Patterns in Indonesia" in the conference *Bangladesh Development Perspectives: Issues in Economic Justice*. We are also thankful to anonymous referees for their invaluable comments and feedbacks. As this article is one of doctoral thesis chapters of the corresponding authors, we would like to thank Dean, School of Professions, supervisors, friends and anonymous examiners in the school of economics at the University of Adelaide, Australia. Usual disclaimer applies.

## Funding:
No fund has been received to prepare this article.